\documentclass[floatfix,reprint,amsmath,amssymb,aps,pra]{revtex4-1}
\usepackage{times,mathptmx}
\usepackage{graphicx}
\usepackage[english]{babel}
\usepackage{epstopdf}
\usepackage{color}

\begin{document}

\title{Mechanism of the ordered particles arrangement
in a concentration grating excited in the field of
counter-propagating Gaussian beams}

\author{A.~A. Afanas'ev}
\author{D.~V. Novitsky}
\email{dvnovitsky@gmail.com}
\affiliation{B.~I.~Stepanov Institute of Physics, National Academy of Sciences of Belarus, Nezavisimosti Avenue 68, 220072 Minsk, Belarus}

\begin{abstract}
In the two-particle approximation, we consider the mechanism for the
formation of an ordered arrangement of transparent spherical
particles of small size in a concentration grating excited by the
gradient force of counter-propagating Gaussian laser beams. This
mechanism is due to the joint action of the transverse gradient
force and the Coulomb force arising as a result of dipole-dipole
interactions between particles.
\end{abstract}

\maketitle

It is known \cite{Smith1981, Rogovin1985, Afanas'ev2005,
Bezryadina2017} that a liquid suspension of transparent particles of
small size is a highly efficient nonlinear optical heterogeneous
medium for continuous laser radiation. For example, as reported in
Ref. \cite{Smith1981}, the optical Kerr coefficient $n_2$ of the
water suspension of latex spherical particles with radius $R=0.117$
$\mu$m and concentration $N=6.5 \cdot 10^{10}$ cm$^{-3}$ turned out
to be $10^5$ times higher than in CS$_2$. For the first time, the
possibility of using such heterogeneous media as a nonlinear optical
material was noted in Ref. \cite{Palmer1980}. Nonlinear optical
phenomena in such artificially created media, in particular,
four-wave mixing \cite{Smith1981, Rogovin1985, Afanas'ev2005} and
stimulated concentration scattering \cite{Afanas'ev2007,
Burkhanov2016}, are caused by energy exchange processes between the
waves interacting on the concentration gratings excited by them due
to the gradient component of the light pressure force.

In Refs. \cite{Burns1989, Afanas'ev2002, Rubinov2003}, the
concentration gratings were recorded in an aqueous suspension of
polystyrene particles with $R = 0.7 –- 3$ $\mu$m with the
interference field of continuous lasers. An interesting experiment
on the excitation of a two-dimensional concentration grating
(matrix) of polystyrene particles with $R = 150 -– 300$ nm in water
under the influence of orthogonal pairs of interfering beams of
continuous laser radiation was performed in Ref. \cite{Mellor2006}.
The results obtained in \cite{Mellor2006} open up the prospects of
synthesizing crystal matrices that demonstrate all the
characteristics of ordinary molecular crystals and are of interest
for various scientific and practical applications.

In the experiments on excitation of concentration gratings by
coherent beams of a He-Ne laser, an ordered arrangement of particles
along its lines was observed \cite{Afanas'ev2002, Rubinov2003}.
Particles ``collected'' under the action of the longitudinal
component of the gradient force $\overrightarrow{F}_z$ in the maxima
of the interference pattern of the field (lines of the concentration
grating) were located at approximately equal distances from each
other (see Fig. 1). For Gaussian beams used in these
experiments, the radiation intensity along the concentration grating
lines is inhomogeneous, so that the transverse component of the
gradient force $\overrightarrow{F}_r$ appears leading to translation
of the particles in the grating. In addition, the Coulomb forces of
the dipole-dipole interaction $\overrightarrow{F}_{dip}$ occur due
to the induced dipole moment between the particles. These forces,
together with the force $\overrightarrow{F}_z$, move the particles
to certain stable positions in the concentration grating.

\begin{figure}[t!]
\includegraphics[scale=1.25, clip=]{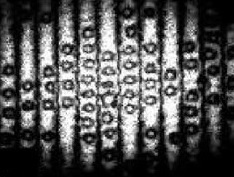}
\caption{\label{fig1} Spatial distribution of polystyrene particles with $R = 2.9$ $\mu$m in the lines of the concentration grating excited by the interference field of He-Ne laser radiation \cite{Afanas'ev2002, Rubinov2003}.}
\end{figure}

In this report, using the two-particle dipole-dipole interaction as
an example, the joint action of the $\overrightarrow{F}_r$ and
$\overrightarrow{F}_{dip}$ forces is considered resulting in the
stable arrangement of the particles in the concentration grating
with a certain distance between them. Analysis of the spatial
distribution of particles under the action of these forces in the
general case is a very complex many-body problem. Nevertheless, even
consideration of two particles gives an idea about the mechanism of
their ordered arrangement in the concentration grating and allows
one to draw certain conclusions and numerical estimates. Although we focus on the Gaussian beams in our consideration, one can find in the literature the other realizations of optical forces using, e.g., the Airy beams \cite{Xie2018a,Xie2018b}.

\begin{figure*}[t!]
\includegraphics[scale=0.75, clip=]{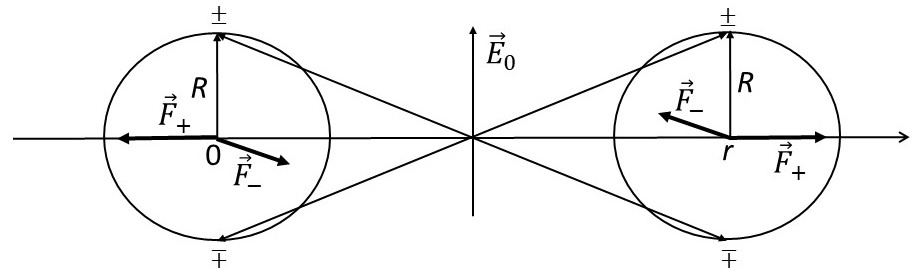}
\caption{Geometric scheme for calculating the Coulomb forces in the interaction of two dipoles induced by the linearly polarized field ${E}_0$.}
\label{fig2}
\end{figure*}

Let us consider a mechanism leading to an ordered arrangement of
transparent spherical particles of small sizes in the concentration
grating excited by linearly polarized Gaussian beams oppositely
propagating along the $z$ axis. In this case, the total amplitude of
the field can be represented as
\begin{equation}
E = E_0 e^{-r^2/2r_0^2} \cos kz \cdot e^{-i \omega t} + c.c.,
\label{eq1}
\end{equation}
where $r_0$ is the beam radius, $r=\sqrt{x^2+y^2}$ is the radial
coordinate. In the field (\ref{eq1}), the particle is affected by
the gradient force given by \cite{Rogovin1985}
\begin{equation}
\overrightarrow{F}_\nabla = \frac{n^2}{2} \alpha \nabla \langle E^2
\rangle_t, \label{eq2}
\end{equation}
where
\begin{equation}
\alpha = \frac{\overline{m}^2-1}{\overline{m}^2+2} R^3 = \alpha_0
R^3, \label{eq3}
\end{equation}
is the particle polarizability ($kR \ll 1$), $\overline{m}=n_0/n$ is
the ratio of the refractive indices of the particle material $n_0$
and the surrounding liquid $n$,
\begin{equation}
\langle E^2 (z,r) \rangle_t = E_0^2 e^{-r^2/r_0^2} (1 + \cos 2kz).
\label{eq4}
\end{equation}
Here $\langle ... \rangle_t$ stands for averaging over time. Further we also assume that the particles are small in comparison to the beam radius, $d=2 R \ll r_0$.

The longitudinal component of the gradient force,
\begin{equation}
F_z = -n^2 \frac{\pi}{\Lambda} \alpha E_0^2 e^{-r^2/r_0^2} \sin 2
\pi \frac{z}{\Lambda}, \label{eq5}
\end{equation}
forms the concentration grating with the period $\Lambda=\pi/k$,
whereas the transversal one,
\begin{equation}
F_r = -n^2 \frac{r}{r_0^2} \alpha E_0^2 e^{-r^2/r_0^2} \left( 1 +
\cos 2 \pi \frac{z}{\Lambda} \right), \label{eq6}
\end{equation}
together with the Coulomb dipole-dipole force $F_{dip}$ results in
the ordered arrangement of particles in the grating (at $z=m
\Lambda$, $m=0, \pm 1, \pm 2...$).

The dipole moment of the particle induced by the field can be
written as follows,
\begin{equation}
p (r,t) = q(r) d \cdot \cos \omega t, \label{eq7}
\end{equation}
where $q(r) = \frac{2 \alpha E_0}{d} e^{-r^2/2 r_0^2}$ is the dipole
charge.

To simplify calculations, we assume that the first particle is
located at the beams center ($r=0$), whereas the second particle is
in the arbitrary point $r$ (see Fig. 2). Since these
coordinates correspond to the particles centers, then $r \gtrsim d$.
Using the geometrical scheme shown in Fig. 2 and the results of Ref. \cite{Rubinov2005}, one can
obtain for our case the expressions for the Coulomb forces of repulsion ($F_+>0$) and attraction ($F_-<0$),
\begin{eqnarray}
F_+ = 8 \left( \frac{\alpha E_0}{rd} \right)^2 e^{-r^2/2 r_0^2},
\nonumber \\
F_-= -\frac{F_+}{(1+d^2/r^2)^{3/2}}. \label{eq8}
\end{eqnarray}
Thus, the resulting Coulomb force is
\begin{equation}
F_{dip} = F_+ + F_- = 8 \left( \frac{\alpha E_0}{rd} \right)^2
e^{-r^2/2 r_0^2} \left[ 1-\frac{1}{(1+d^2/r^2)^{3/2}} \right] >0.
\label{eq9}
\end{equation}
From Eq. (\ref{eq6}) at $z=m \Lambda$ and Eq. (\ref{eq9}), we have
the transcendental equation for the equilibrium distance $r_p$
between the particles,
\begin{equation}
\frac{\alpha_0 d r_0^2}{2 n^2 r_p^3} \left[
1-\frac{1}{(1+d^2/r_p^2)^{3/2}} \right] e^{r_p^2/2 r_0^2} =1.
\label{eq10}
\end{equation}
Obviously, the distance $r_p$ does not depend on the radiation
intensity. It is determined by the polarizability and particle size
as well as the radius of laser beams. In particular, for $r_p \ll
r_0$ and $(d/r_p)^2 \ll 1$, we find from Eq. (\ref{eq10}) that
\begin{equation}
r_p \approx \sqrt[5]{6 \alpha_0 (r_0/n)^2}. \label{eq11}
\end{equation}
For example, for polystyrene particles with $R=100$ nm in water
($n=1.33$, $\alpha_0=0.12$) and beams radius $r_0=4 \cdot 10^{-2}$
mm, we obtain $r_p \approx 0.9 \cdot 10^{-3}$ mm. At the same time,
both inequalities determining the range of applicability of Eq.
(\ref{eq11}) are satisfied. With an increase in the particle radius
by $2$ times, the distance $r_p$ increases by about $1.5$ times.

In conclusion, we note that in the experiments \cite{Afanas'ev2002,
Rubinov2003}, the small-particle approximation ($kR \ll 1$) used
here is not fulfilled and, accordingly, the formalism of the Maxwell
stress tensor \cite{Stratton} must be used to calculate the gradient
force. This approach allows us to calculate the gradient force in
the case of particles of arbitrary radius \cite{Guzatov2008}.
However, such calculations are very time consuming and require
numerical simulations. For this brief report, which explains only
the mechanism of the ordered arrangement of particles in the
concentration gratings, these cumbersome calculations are not given
here. We also note that when a concentration grating is excited by
plane waves, the arrangement of the particles in its lines will be
determined by the Coulomb forces of the dipole-dipole interaction
and the transverse dimensions of the cell containing the suspension. The considered mechanism based on competition of the gradient force and dipole-dipole interaction should be taken into account upon creation of molecular-crystal-like structures \cite{Mellor2006} from liquid suspensions of small particles with the optical methods.

The authors are grateful to Prof. A.N. Rubinov for useful
discussions.

\end{document}